\documentclass[prl,twocolumn]{revtex4}
\usepackage{graphicx,amsmath,amssymb}
\renewcommand{\narrowtext}{\begin{multicols}{2} \global\columnwidth20.5pc}
\renewcommand{\v}[1]{{\bf #1}}

\def\be{\begin{eqnarray}}
\def\ee{\end{eqnarray}}
\newcommand{\nn}{\nonumber\\}
\newcommand{\Eq}[1]{Eq.~(\ref{#1})}
\newcommand{\Fig}[1]{Fig.~(\ref{#1})}

\begin{document}
\draft

\title{Incompressible Quantum Liquids and New Conservation Laws}

\author{Alexander Seidel$^{a,b}$, Henry Fu$^{a}$, Dung-Hai Lee$^{a,b,c}$,
Jon Magne Leinaas$^d$ and Joel Moore$^{a,b}$}

\affiliation{${(a)}$Department of Physics,University of California
at Berkeley, Berkeley, CA 94720, USA}
 \affiliation{${(b)}$
Material Science Division, Lawrence Berkeley National
Laboratory,Berkeley, CA 94720, USA.}
 \affiliation{${(c)}$ Center
for Advanced Study, Tsinghua University, Beijing 100084, China.}
\affiliation{${(d)}$ Department of Physics,University of Oslo,
P.O. Box 1048 Blindern, 0316 Oslo, Norway. }

\date{\today}

\begin{abstract}

In this letter we investigate a class of Hamiltonians which, in
addition to the usual center-of-mass (CM) momentum conservation,
also have center-of-mass position conservation.  We find that
regardless of the particle statistics, 
the energy spectrum is at least q-fold
degenerate when the filling factor is $p/q$, where $p$ and $q$
are coprime integers. Interestingly the
simplest Hamiltonian respecting this type of symmetry encapsulates
two prominent examples of novel states of matter, namely the
fractional quantum Hall liquid and the quantum dimer liquid.  We
discuss the relevance of this class of Hamiltonian to the search
for featureless Mott insulators.


\end{abstract}

\maketitle

In the twentieth century the basic notions of ``symmetry'' and
``order'' have set a paradigm that gave rise to a lasting
cross-fertilization between various branches of physics.
Landau\cite{landau} first put forth the idea that in the process
of ordering a system also becomes less symmetric. This phenomenon
is commonly known as ``symmetry breaking'', and underlies our
general understanding of order in a wide variety of systems.
Ordering is such a ubiquitous tendency of nature that it is
difficult to give an example where a system remains unordered to
the lowest temperatures.
In recent years condensed matter physicists have been asking
whether the electrons in a solid can remain unordered at absolute
zero temperature.

Following Anderson's proposal of the ``spin
liquid''\cite{anderson}, a novel magnetic state with no order, for
the past twenty years condensed matter physicists have expended
tremendous effort searching for unordered electronic states in
solids.\cite{wen,wen2,wenbook}. Recently there has been
significant theoretical progress in demonstrating the stability of
some examples of such states, and hence proving that they are in
principle possible\cite{z2,rantner,hermele}. In addition, it has
been shown that there is a strong tie between such states and the
phenomenon of ``quantum number fractionalization''.\cite{senthilosh}
Concurrent with these efforts there has been an
extensive pursuit for microscopic models exhibiting these
properties.\cite{rk,ms,Lhuillier,balents,misguich,senthil} The
traditional guideline  for such pursuits has been ``frustration''
(see, e.g. \cite{fezekas,Subir,Moessner}).
 In this letter we adopt a different
viewpoint, and investigate whether quantum disordered fermion or
boson states  can arise from unusual conservation laws.


It is commonly perceived that one of the most difficult situations
for a system to be free of symmetry breaking is in fractionally
filled
Mott insulators. For bosons, the filling
factor is just the average number of particles per crystalline unit
cell, and for electrons it is half that
quantity. 
In this paper, we view any non-disordered insulating state
as a Mott insulator if the insulating property cannot be attributed
to the Pauli exclusion principle.
  Whether a fractionally filled Mott insulator can be
''featureless" ({\it i.e. unordered}) has been a central issue of
debate in recent years. 
 So far all known such systems possess order\cite{1dex}.
 For example, a half-filled square lattice of boson
atoms is only known to be Mott insulating when the atoms localize
in a checkerboard pattern 
and hence break the symmetry of the
original lattice \cite{bosonMott}. 


Physicists have only recently begun to understand the reason why
symmetry breaking usually accompanies Mott insulators. 
This progress is due to generalizations\cite{oshikawa, hastings}
of an earlier theorem by Lieb, Schultz and
Mattis.\cite{lsm} In particular, Oshikawa\cite{oshikawa, note} has 
argued that when a 
fractionally filled
Mott insulator has an energy gap between its ground state(s) and
excited states, it must have ground state
degeneracy. In particular, if the filling factor is
$p/q$, the ground state must be at least $q$-fold degenerate.
Empirically, this degeneracy is always achieved by breaking
symmetry. In the boson Mott insulator example given above, a
2-fold degeneracy arises because there are two distinct,
complementary, checkerboard-ordered patterns. For the time being
we will use the phrase ``Mott insulator'' to refer to a
fractionally filled Mott insulator with an energy gap.

Symmetry breaking is one way to produce the degeneracy required by
Oshikawa's result. However if there were a way to produce this
degeneracy independently of symmetry breaking, it would provide a
candidate route to finding featureless Mott insulators. In the
following we show that imposing center-of-mass (CM) momentum and
position conservation is one such way.
\smallskip
\\
{\noindent{{\bf Theorem: Simultaneous conservation of
center-of-mass momentum and position
guarantees q-fold degeneracy of the energy
spectrum.}}}
Consider a D-dimensional lattice. 
We impose periodic boundary conditions in a certain direction, say
the x-direction, of the lattice. Let L be the spatial period in
this direction, and  $T$ be the operator of translation by one
lattice constant in the x-direction.  The ``cross section'' $C$ of
the lattice is defined such that the total number of unit cells is
given by $CL$. Following Oshikawa, let us now consider filling
factor $\nu=p/q$, and $C$ relatively prime to $q$. In this case
Oshikawa's argument tells us that the ground state is at least
q-fold degenerate if there is an energy gap.

The (exponentiated) CM position in the x-direction modulo $L$ is given by 
\vspace{-.15cm}
\be
U=\exp{\Big(i{ 2\pi\over L} \sum_{\vec{r}}
x~C^\dagger_{\vec{r}}C_{\vec{r}}\Big)}. \ee 
The simultaneous
conservation of the CM momentum and position implies $[H,U]=
[H,T]=0.$  Because of this, we can choose the eigenstates of $H$
to be simultaneous eigenstates of $U$. In addition since 
\vspace{-.15cm}
\be
T^{-1}~U~T=\exp{\Big[i{ 2\pi\over L} \sum_{\vec{r}}
(x+1)~C^\dagger_{\vec{r}}C_{\vec{r}}\Big]}=e^{i2\pi C {\frac{p}
{q}}}~U,\ee  \vspace{-.2cm}
we obtain
 \vspace{-.0cm}
\be
U~T= e^{i2\pi C {\frac{p}{q}}}~TU.\label{Stu} 
\ee
Now we show that $q$ consecutive actions of $T$ on any energy
eigenstate generates q degenerate orthogonal states differentiated
by their eigenvalue with respect to $U$. Let the eigenstate in
question be denoted by $|e^{i\phi},E>$ where $e^{i\phi}$ is the
eigenvalue with respect to $U$ and $E$ is the energy eigenvalue.
Because the translation operator commutes with the Hamiltonian,
the state $T|e^{i\phi},E>$ is also an eigenstate of $H$ with the
same energy. However its eigenvalue with respect to $U$ is \be
U~T|e^{i\phi},E> &&=e^{i2\pi C {p\over q}}~TU|e^{i\phi},E>\nn&&
=\Big(e^{i2\pi C {p\over q}}e^{i\phi}\Big)~T|e^{i\phi},E>.\ee The
different eigenvalue with respect to $U$ implies that the two
energy eigenstates $|e^{i\phi},E>$ and $T|e^{i\phi},E>$ are
orthogonal. We can perform this operation $q$ times before we
arrive at a state with the original $U$ eigenvalue. Hence there is
at least a $q$-fold degeneracy.

Motivated by the above observations, we would like to analyze
Hamiltonians featuring CM position and momentum conservation to
determine if they lead to featureless Mott insulators.
 The simplest non-trivial such Hamiltonian is 
\vspace{-.15cm}
\be
 \qquad H= \sum_{\v R,\v x,\v y}
 g(\v x,\v y)~C^\dagger_{\v R+\v x}C^\dagger_{\v R-\v x}~C_{\v R-\v y}C_{\v R+\v y}.\label{Spairhopp}
\ee 
This Hamiltonian hops a pair of particles while conserving
their CM position. In general it is quite difficult to solve for the
eigenstates of this type of Hamiltonian. However, for a special
one-dimensional case some insight can be gained due to a
connection with a well-studied problem in physics, the fractional
quantum Hall effect. 
To be specific, we consider the the Trugman-Kivelson (TK) Hamiltonian\cite{tk}, which has Laughlin's 1/3 wavefunction\cite{laughlin} as the ground state. When studied on a torus with dimensions $L_x$ and $L_y$, the TK Hamiltonian can be written in terms of the creation/annihilation operators of the lowest Landau level as follows\cite{ll}:
\be
&&H=\sum_{R,x,y} f^*(x)f(y) C^\dagger_{R+x}C^\dagger_{R-
x}~C_{R-y}C_{R+y},\nn &&f(x)=\kappa^{3/2}\sum_{n} (x-nL)
e^{-\kappa^2(x-nL)^2} \label{pairhopp}. \ee Here \be L=L_xL_y/2\pi
l_B^2,~~~ \kappa=2\pi l_B/L_y.\label{redef} \ee In the above
$l_B=\sqrt{\hbar c/eB}$ is the magnetic length.  This Hamiltonian
acts upon a system of spinless fermions on a ring of $L$ sites at
filling factor 1/3.

One important consequence of the connection between \Eq{pairhopp}
and the quantum Hall effect on a torus is an unusual property of
\Eq{pairhopp}, which we call duality. By performing Fourier
transforms it is simple to show that when written in terms of
operators that create and annihilate particles in a fixed momentum
state, \Eq{pairhopp} reads \be
 H= \sum_{Q,k,q} \tilde{f}^*(k)\tilde{f}(q)~C^\dagger_{Q+k}C^\dagger_{Q-k}~C_{Q-q}C_{Q+q},\label{pairhoppp}
\ee where $\tilde{f}(k)=\sum_x e^{i2kx}f(x)$. For the $f$ used in
\Eq{pairhopp} $\tilde{f}$ has the same form as $f$. This implies
that the Hamiltonian in the momentum space is also described by
\Eq{pairhopp} except $\kappa\rightarrow {2\pi\over \kappa L}$. The
factor $2\pi/L$ arises from the lattice constant in the reciprocal
space. For each $L$ there is a special $\kappa$ value
($\kappa^*=\sqrt{2\pi/L}$) for which the real space and momentum
space $\kappa$ are the same.  The duality implies that energy
spectra at $\kappa$ and $2\pi/\kappa L$ are identical. From the
perspective of the quantum Hall liquid on a torus, the duality
merely signifies that interchanging $L_x$ and $L_y$ leads to the
same physical system.

Analyzing the  Hamiltonian \Eq{pairhopp} we find that for all
practical purposes the ground state is a featureless Mott
insulator when $\kappa$ is small. In addition, we find the
surprising fact that  on a torus with finite circumference in one
direction (but infinite circumference in the other) the Laughlin
liquid has non-vanishing density wave order, and is {\it
adiabatically connected} to a Wigner crystal.

First we focus on the large $\kappa$ limit. In the large $\kappa$
limit we can expand the Hamiltonian in the
 parameter $e^{-\kappa^2}$. As a result
\Eq{pairhopp} is accurately approximated by \be
H=\sum_{i=1}^L\Big[ f(1/2)^2 n_{i+1}n_i + f(1)^2
n_{i+2}n_{i}\Big], \label{Spairhop1} \ee where $n_i=C^+_iC_i$.
This Hamiltonian imposes energy penalties for having nearest
neighbor and next-nearest neighbor particles. Its ground states
are "Wigner crystals" where one out of every three lattice sites is
occupied. Clearly at $\nu=1/3$ there are three such ground states.
These ground  states possess density wave order (hence are not
featureless).

\begin{figure}
\includegraphics[width=6.5cm]{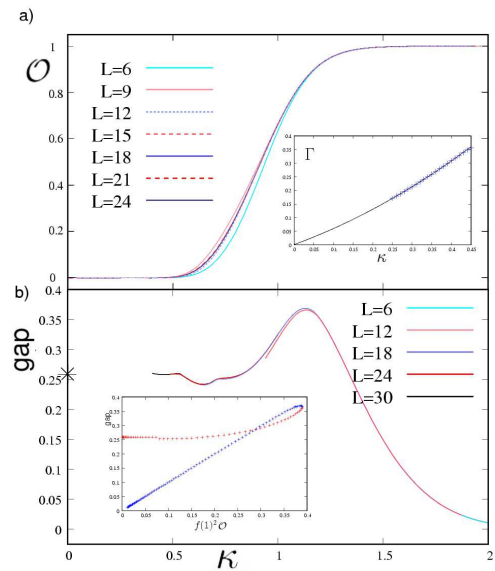}
\caption {The density wave order parameter and the energy gap of
\Eq{pairhopp} as a function of $\kappa$. Data are obtained for
$L\leq 30$. a) The crystalline order parameter. Inset: The
quantity $\Gamma=(-\log{\cal O})^{-1/2}$, extrapolated to
$L=\infty$ using the alternating epsilon algorithm.\cite{barber}
The smooth curve is a polynomial fit of the data for
$.25<\kappa<.45$ (crosses). b) The energy gap for data restricted
to $\kappa L >12$ and even particle number (see text).
Inset: The energy gap is plotted versus $ {\cal O} f(1)^2$. Blue
and red symbols are for $\kappa
>1.1$ and $\kappa <1.1$ respectively. Two distinct branches are
clearly visible. 
}\label{data}
\end{figure}

To quantify the degree of crystalline order we introduce the order
parameter ${\cal O}$
associated with a spatial period of three lattice constants:
\begin{equation}
  {\cal O}=\frac{1}{N} \sum_{j=1}^L e^{i{2\pi\over 3}j}\, <n_j>.
\end{equation}
In the above $N$ is the total particle number, and $<n_j>$ is the
expectation value of the site occupation in (any one of) the
ground state(s). ${\cal O}$ is normalized such that its modulus
becomes unity in the extreme crystalline limit.
 In \Fig{data}a) we plot ${\cal O}$
as a function of $\kappa$, evaluated numerically for system sizes
$L=6,9,...,24$. For $\kappa\geq 1.5$ we obtain ${\cal{O}}\approx
1$ implying that the ground state is nearly a perfect crystal.  As
expected, as $\kappa$ decreases the order parameter decreases and
practically vanishes below $\kappa\approx 0.5$.\cite{haldane} This
finding raises an important question: Is the regime below
$\kappa=.5$ a featureless Mott insulator? The weak size dependence
of the crystalline order parameter near $\kappa=0.5$ is
inconsistent with the existence of a continuous phase transition,
which is required if the state below $\kappa=0.5$ is truly free of
crystalline order. The smooth evolution of ${\cal O}$ with
$\kappa$ rules out the possibility of a first-order phase
transition. Indeed, while the order parameter becomes
exponentially small at small $\kappa$ (numerical precision limits
our study to $\kappa\gtrsim .25$.), a careful analysis of the
numerical data indicates that at any finite $\kappa$, it will not
vanish as $L\rightarrow\infty$.  Using standard methods of
extrapolation \cite{barber} we find that the order parameter
for $L\rightarrow\infty$ is given by $\exp(-1/\Gamma(\kappa)^2)$,
where $\Gamma(\kappa)$ vanishes linearly at $\kappa=0$ (see inset
in \Fig{data}(a)).  According to this result, density wave order
exists as long as $\kappa\ne 0$.

In terms of the quantum Hall connection the fact that the Laughlin
liquid on a torus with any finite $L_y$ (and infinite $L_x$) has a
nonzero density wave order is very surprising (to us at least).
From the perspective of a 1D lattice problem, however, it can be shown
that in general there exists a local order parameter that
can be used to distinguish the degenerate ground states required
by Oshikawa's argument.\cite{affleck, tasaki}
However, e.g., for $\kappa=0.25$ the crystal order is 15 orders of
magnitude weaker than that at $\kappa=1.5$! Hence for all
practical purposes this is a state without order.

In the following we carefully study the energy gap as a function
of $\kappa$ to demonstrate that 1) the system is a Mott insulator
for all $\kappa$, and 2) there is no quantum phase transition
between the large and small $\kappa$ regimes.  In \Fig{data}b) we
show the numerically computed energy gap vs. $\kappa$ for system
sizes up to $L=30$. Only data for even particle number (to avoid
even/odd effect) are used. The limit $L\rightarrow\infty$ and
$\kappa$ fixed amounts to studying a Hall torus with
$L_x\rightarrow\infty$ and fixed $L_y$ fixed. To avoid finite size
effect from small $L_x$ we restrict $L_x>12 l_B$ or equivalently,
$\kappa L >12$ (for $L=30$, $\kappa>0.4$). With these restrictions
the data points collapse onto a single curve, as shown in
\Fig{data}(b). Note that according to this result, the gap remains
robust even in the ``transition regime'' around $\kappa=0.5$.
According to quantum Hall physics the energy gap should not depend
on $L_x$ or $L_y$ when both are much greater than $l_B$. Thus for
$\kappa<0.4$ (or $L_y> 5\pi l_B$ for the Hall torus), we expect
little variation in the extrapolated gap value. The asterisk in
\Fig{data}(b) is obtained by extrapolating the gap value at
$\kappa^*(L)$ to $L=\infty$ (note that
$\lim_{L\rightarrow\infty}\kappa^*(L)=0$).

From these results the energy gap is nonzero for all $\kappa$.\cite{note2}
However, in 
the inset of 
\Fig{data}b) we demonstrate that two different
mechanisms cause the energy gap at small and large $\kappa$. For
large $\kappa$ the gap is the energy penalty of having
next-nearest-neighbor particles. The latter quantity is
proportional to $f(1)^2{\cal O}$.  However the energy gap plotted
against $f(1)^2{\cal O}$ exhibits two branches, indicating a
different mechanism for the energy gap at small $\kappa$ values.
Hence the small $\kappa$ state is not just an ordinary crystal
with extremely small order. Thus even in the most unfavorable case
(1D) , our approach has succeeded in producing a Mott insulator
that is practically featureless. In higher dimensions, where there
are no theorems forbidding featureless states, we expect
\Eq{Spairhopp} will lead to {\em truly} featureless Mott
insulators (see later discussion on quantum dimer models).

Our analysis also leads to important implications for the
fractional quantum Hall liquid on a torus.
In the quantum Hall community, it is well known that when
$L_y$ is large compared to $l_B$ (i.e. when $\kappa$ is small),
the Laughlin state on the torus should be practically
indistinguishable from that in the infinite 2D plane
($L\rightarrow\infty, \kappa=0$). Quite unexpectedly, however, our
 findings imply that the Laughlin state in this regime is
adiabatically connected to a state
with strong crystalline order at small $L_y$.\cite{tth} 
The above result reveals a danger in over-interpreting adiabatic
continuity, because the Laughlin state is clearly qualitatively
different from the electron crystal. However, since discrete
quantum numbers are preserved by adiabatic continuity, both the
ground state degeneracy and the quasiparticle charge are the same
in the quantum Hall and the crystal state. The obvious threefold
degeneracy of the electron crystal originates from the three
distinct center-of-mass positions of the system. The excited
states of the crystal are domain walls carrying $\pm 1/3$ charge
due to the Su-Schrieffer counting argument\cite{sshsu}.

Our findings are consistent with the notion that the Laughlin
state on a finite torus has topological order. As proposed by Niu
and Wen\cite{wen}, unlike in a symmetry broken state, disorder
lifts the ground state degeneracy of a topologically ordered state
by an amount that is exponentially small in the system size. In
our case, such exponential dependence comes from the weak order
parameter itself. Based on this we conclude that despite the
common perception, the presence of symmetry breaking order should
not be taken as excluding topological order.

Our belief that \Eq{Spairhopp} can lead to truly featureless Mott
insulators in higher dimensions is supported by the following
connection to the quantum dimer model.\cite{rk} The quantum dimer
model is a special case of the bosonic version of \Eq{Spairhopp}
in two dimensions. Indeed, once dimers are reinterpreted as point
bosons residing at bond centers, the quantum dimer Hamiltonian
becomes a center-of-mass position conserving model described by
\Eq{Spairhopp}.  Moessner and Sondhi\cite{ms} argued that the
quantum dimer model on a triangular lattice exhibits an unordered
quantum phase.  This supports our  belief that \Eq{Spairhopp}
leads to truly featureless Mott states in higher dimensions.

Finally, although we have focused on Mott insulators, gapless
states described by \Eq{Spairhopp} are equally interesting. We can
prove that such a gapless system cannot be an ordinary metal
(fermions) or a superfluid (bosons), because center-of-mass
position conservation implies the absence of Drude
weight/superfluid density. For example in one space dimension such
a gapless liquid will not fall within the usual Luttinger liquid
paradigm.

To conclude we have presented the idea that simultaneous
conservation of CM position and momentum can lead to fractionally
filled featureless Mott insulators. It is remarkable that the
simple model given by \Eq{Spairhopp} unifies systems as diverse as
the quantum Hall liquid and the quantum dimer liquid, whose effective field
theories are as different as Chern-Simons gauge theory and Z2
gauge theory, respectively. 
We hope our results will spur the exploration
of new directions and lead to a wealth of new states of matter
that have, so far, escaped our attention.

Acknowledgement: DHL and AS are supported by DOE grant
DE-AC03-76SF00098. HCF was partially supported by a predoctoral
fellowship from the Advanced Light Source.
\vspace{0.1in}

\begin{thebibliography}{99}
\bibitem{1dex} 
There is a special example of a featureless Mott insulator,
namely, the 1D Hubbard insulator at half filling; however this
system is gapless due to the spin degrees of freedom. In the following
we refer to fully gapped systems, see below. In
addition, the even-leg Hubbard ladders have often been given as an
example of fully gapped featureless Mott insulators; however they
have an integer filling factor.
\bibitem{note}
Strictly speaking, Oshikawa's result was formulated only for
spinless particles. Indirectly, however, it also applies to the
interesting case of Mott insulating electron systems with filling
factor $1/2$. In this case the physics is described by local
spin-$\frac{1}{2}$ degrees of freedom. These spins can be
represented as a hardcore boson system at filling factor $1/2$, for
which Oshikawa's arguments do apply.
\bibitem{note2}
Note that the overall energy scale of the problem,
$f(1)^2$, becomes exponentially small around $\kappa\approx 2$.
\end{thebibliography}
\end{document}